\documentclass[twocolumn,english]{revtex4-1}
\usepackage[T1]{fontenc}
\usepackage[latin9]{inputenc}
\usepackage{amstext}
\usepackage{amssymb}
\usepackage{graphicx}
\usepackage{babel}
\usepackage{hyperref}
\begin{document}
\title{Dynamic magnetism in the disordered hexagonal double perovskite BaTi$_{1/2}$Mn$_{1/2}$O$_{3}$}
\author{M.R. Cantarino$^{1}$, R.P. Amaral$^{2}$, R.S. Freitas$^{1}$, J.C.R.
Ara\'ujo$^{2}$, R. Lora-Serrano$^{2}$, H. Luetkens$^{3}$, C. Baines$^{3}$,
S. Br\"auninger$^{4}$, V. Grinenko$^{4}$, R. Sarkar$^{4}$, H.H. Klauss$^{4}$,
E. C. Andrade$^{5}$, F. A. Garcia$^{1}$ }

\affiliation{$^{1}$IFUSP, Universidade de S\~ao Paulo, 05508-090, S\~ao Paulo-SP, Brazil}

\affiliation{$^{2}$Universidade Federal de Uberl\^andia, Instituto de F\'isica, 38400-902,
Uberl\^andia-MG, Brazil}

\affiliation{$^{3}$Laboratory for Muon Spin Spectroscopy, Paul Scherrer Institute,
CH-5232 Villigen PSI, Switzerland}

\affiliation{$^{4}$Institute for Solid State and Material Physics, TU Dresden,
D-01069 Dresden, Germany}

\affiliation{$^{5}$Instituto de F\'isica de S\~ao Carlos, Universidade de S\~ao Paulo,
C.P. 369, S\~ao Carlos, SP, 13560-970, Brazil}
\begin{abstract}
Magnetic frustration and disorder are key ingredients to prevent the
onset of magnetic order. In the disordered hexagonal double perovskite
BaTi$_{1/2}$Mn$_{1/2}$O$_{3}$, Mn$^{4+}$ cations, with $S=3/2$
spins, can either form highly correlated states of magnetic trimers
or dimers or remain as weakly interacting orphan spins. At low temperature
($T$), the dimer response is negligible, and magnetism is dominated
by the trimers and orphans. To explore the role of magnetic frustration,
disorder and possibly of quantum fluctuations, the low-$T$ magnetic
properties of the remaining magnetic degrees of freedom of BaTi$_{1/2}$Mn$_{1/2}$O$_{3}$
are investigated. Heat-capacity data and magnetic susceptibility display
no evidence for a phase transition to a magnetically ordered phase
but indicate the formation of a correlated spin state. The low-temperature
spin dynamics of this state is then explored by $\mu$SR experiments.
The zero field $\mu^{+}$ relaxation rate data show no static magnetism
down to $T=19$ mK and longitudinal field experiments support as well
that dynamic magnetism persists at low $T$. Our results are interpreted
in terms of a spin glass state which stems from a disordered lattice
of orphans spins and trimers. A spin liquid state in BaTi$_{1/2}$Mn$_{1/2}$O$_{3}$, however, is not excluded and is also discussed. 
\end{abstract}
\maketitle

\section{Introduction}

Transition metal ($M$) oxides are forever the source of intriguing
physics in the field of correlated electron systems, including a large
number of complex magnetic systems \cite{dagotto_complexity_2005}.
Complexity emerges from the magnetic interaction between the spins
at the $M$ sites, which takes place by means of a superexchange interaction
\cite{goodenough_magnetism_1963} that will usually drive the system
to a magnetic ordered state \cite{white_quantum_2007}. In some materials,
however, dimensionality, disorder, lattice geometry or the symmetry
of the interactions may hinder the appearance of long range order
\cite{anderson_resonating_1973,read89,kitaev_anyons_2006,helton07,mourigal13}.
In these cases, the system will remain in a dynamic magnetic state.
This dynamic magnetic state may either be identified as a spin freezing
phenomenon \cite{uemura_muon_1985,moessner_geometrical_2006} or,
at sufficient low temperatures $(T)$, the system may be driven by
quantum fluctuations to some exotic ground state, such as a quantum
spin liquid ground state. A spin liquid state is a disordered quantum
magnetic state wherein the spins are entangled in a highly correlated
and dynamic state of matter \cite{balents_spin_2010,savary17,zhou_quantum_2017}. 

The first task to experimentally verify the formation of a spin liquid
state is to provide evidence for the lack of phase transitions down
to the lowest temperatures. It is paramount, as well, to present evidence
that a collective spin state is formed at low-$T$, most striking
by characterizing continuous magnetic excitations \cite{han_fractionalized_2012,paddison_continuous_2017,balz_physical_2016}
or the spin relaxation rate regime. For the later purpose, muon spin
relaxation ($\mu SR$) is a particular appropriate technique \cite{uemura_spin_1994,blundell_spin-polarized_1999,li_muon_2016,khuntia_spin_2016}.
The task, however, cannot be settled based upon $\mu SR$ measurements
alone, since a dynamic magnetic state without strong quantum fluctuations
(such as a spin glass) will also display an evolution of the relaxation
regime \cite{uemura_muon_1985,tran_low-temperature_2018} which hold
some similarity to what is found for some spin liquid candidates \cite{khuntia_spin_2016,ding_possible_2018}. 

In the disordered hexagonal double perovskite BaTi$_{1/2}$Mn$_{1/2}$O$_{3},$
$S=3/2$ spins due to Mn$^{4+}$cations pair up to form magnetic dimers
in a singlet state, coexisting with orphan spins and magnetic trimers
\cite{garcia_magnetic_2015} . The formation of dimers and trimers
are dictated by large energy scales ($\approx101-176$ K). The effective
low-$T$ degrees of freedom involve orphan spins and trimers, distributed
in layers of triangular lattices, with competing antiferromagnetic
exchange interactions. The distinct layers are well spaced, making
the interactions mainly of intralayer character. As we shall discuss,
this distribution of orphans and trimers implies a large magnetic
frustration, with a lower bound for the frustration parameter ($f$)
of about $\approx75$. 

In this paper, we show that this configuration paves the way for the
existence of a dynamic magnetic state at low $T$ in our system, comprised
of correlated $S=3/2$ (orphans) and effective $S=1/2$ spins (trimers).
We present and discuss low-$T$ heat capacity and susceptibility data
which show no sign of long-range order, but which indicate a highly
correlated magnetic state. Subsequently, we explore the spin dynamics
of our system by $\mu SR$ , and it is shown that dynamic magnetism
persists down to the lowest temperatures ($T=0.019$ K) achieved in
our experiments. Our results and analysis support that this dynamic
magnetic state can be interpreted as a spin glass state. However,
a spin liquid scenario is not excluded and is discussed as well. 

\section{Methods}

High quality polycrystalline samples were synthesized by the solid
state reaction method as in Ref. \cite{garcia_magnetic_2015}. Synchrotron
X-ray diffraction experiments were carried out at the XRD$1$ beamline
of the LNLS-CNPEM, Brazil \cite{carvalho_x-ray_2016}. Heat capacity
measurements were performed to temperatures down to $0.1$ K and magnetic
fields up to $9$ T, employing a Quantum Design PPMS, using a dilution
refrigerator. DC and AC susceptibility ($\chi(T)$ and $\chi'(T)$)
measurements were performed for temperatures down to $0.6$ K employing
a Quantum Design MPMS ($1.8-300$ K) and a Vibrating Sample Magnetometer
(VSM) in a $^{3}$He cryostat. 

$\mu$SR measurements have been carried out at the $\pi$M$3$ beam
line at the GPS and LTF spectrometers of the Swiss Muon Source at
the Paul Scherrer Institute (PSI) in Villigen. The measurements were
performed in the $T$ interval $0.019<T<10$ K (LTF) and $1.5<T<200$
K (GPS) \cite{amato_new_2017} in zero magnetic field (ZF) and in
longitudinal applied magnetic fields (LF) with respect to the initial
muon spin polarization up to $0.05$ T. To improve the thermal contact,
the samples in the experiments at the LTF spectrometer were glued
on a Ag plate. This gives rise to a time and temperature independent
background due to muons that stopped in the Ag plate. The $\mu$SR
time spectra were analyzed using the free software package MUSRFIT
\cite{suter_musrfit:_2012}. DFT calculations of the electrostatic
potential were performed to extract information about the muon local
environment. We employed the all-electron full-potential linearized
augmented-plane wave code Elk 4.3.6, which includes the Spacegroup
package using Broyden mixing \cite{srivastava_broydens_1984} and
the Generalized Gradient Approximation functional \cite{perdew_generalized_1996}.

\section{Results and Discussion}

\subsection{Structure, magnetism and heat capacity}

\begin{figure}
\begin{centering}
\includegraphics[scale=0.3]{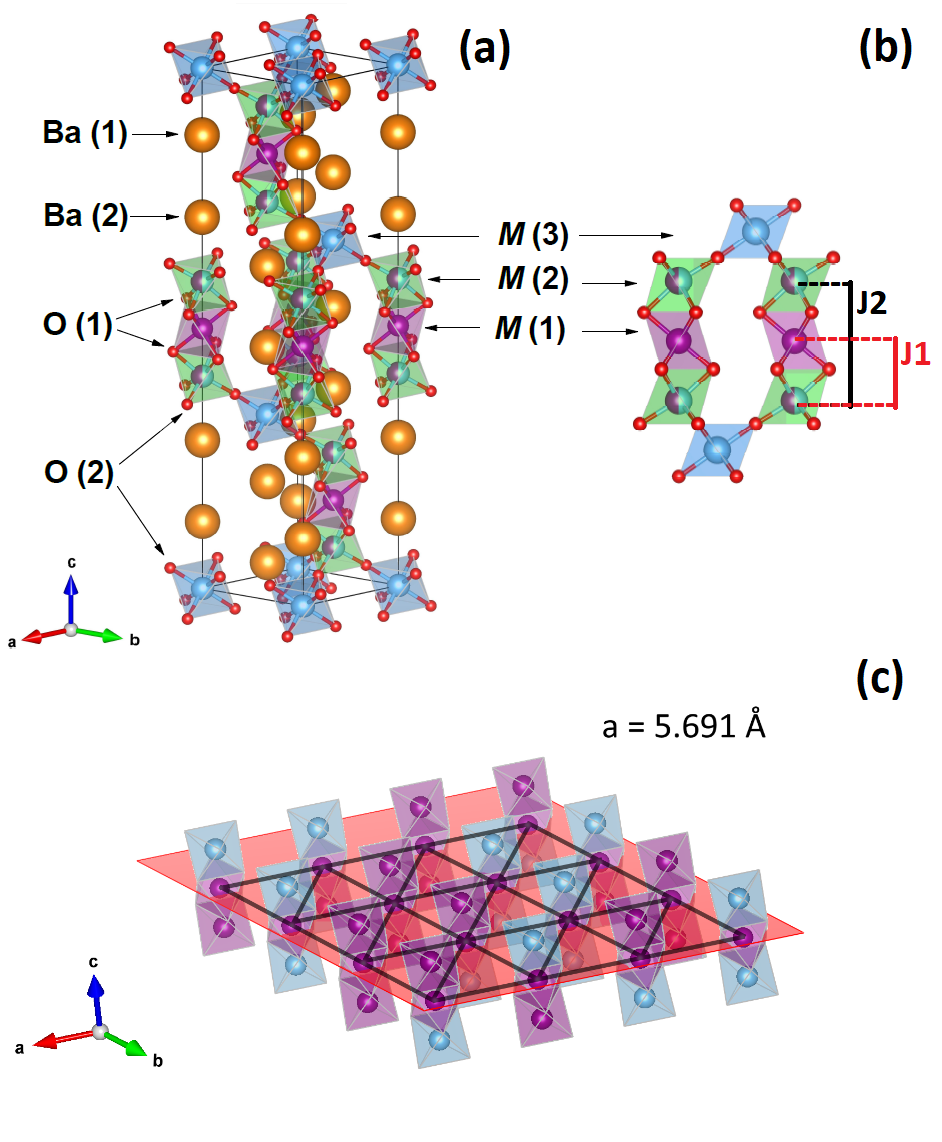}
\par\end{centering}
\caption{$(a)$ Structural model of BaTi$_{1/2}$Mn$_{1/2}$O$_{3}$ with a
12R-type perovskite structure (as in Ref. \cite{garcia_magnetic_2015}).
Barium (Ba$(1)$ and Ba$(2)$), oxygen (O$(1)$ and O$(2)$) and transition
metal ($M(1)$, $M(2)$ and $M(3)$) sites are indicated. $(b)$ Detail
of the structural trimer indicating the $J_{1}$ and $J_{2}$ exchange
constants. $(c)$ Model of the proposed emerging magnetic lattice
at low $T$. Half of the $M(2)$ sites are filled with Mn atoms (purple)
and the other half by Ti atoms (blue), in a disordered way, that is
statistically determined. \label{fig:Structure}}
\end{figure}

The BaTi$_{1/2}$Mn$_{1/2}$O$_{3}$ structure is presented in Fig.
\ref{fig:Structure}$(a)$ \cite{garcia_magnetic_2015,keith_synthesis_2004,miranda_compositionstructureproperty_2009}
(see Ref. \cite{Cantarino_supplemental} for further support from
new resonant diffraction experiments). The structure possesses $3$
transition metal sites, termed $M(1)$, $M(2)$ and $M(3)$. Mn cations
are found at the $M(1)$ and $M(2)$ sites, inside the structural
trimers. The $M(1)$ sites are occupied exclusively by Mn atoms whereas
the occupation of the $M(2)$ sites is mixed.

All analysis so far \cite{garcia_magnetic_2015,keith_synthesis_2004,miranda_compositionstructureproperty_2009}
and our results \cite{Cantarino_supplemental} support that the $M(2)$
site is equally occupied by either Mn or Ti atoms. Therefore, the
relative fraction of dimers, trimers and orphans are statistically
determined as in Ref. \cite{garcia_magnetic_2015}. It results that
$1/8$ of the Mn are orphans, half pair up to form dimers and $3/8$
form trimers. The coupling of magnetic dimers and trimers are described
by two strong exchange constants $J_{1}$ and $J_{2}$ (Fig. \ref{fig:Structure}$(b)$),
which were shown to be of comparable size \cite{garcia_magnetic_2015}. 

As a function of $T$, the dimer contribution to magnetism is exponentially
suppressed and we can safely assume that at low-$T$ the remaining
magnetic degrees of freedom are due to orphan spins and magnetic trimers.
Focusing on the plane formed by the $M(1)$ sites, one may propose
the following emerging $2$D structure for the magnetic lattice, represented
in Fig. \ref{fig:Structure}$(c)$: layers of randomly distributed
magnetic trimers and orphans and dimers in a triangular lattice (lattice
constant $a=5.69\text{ }\mathring{\text{A}}$). The dimers, being
non-magnetic at low $T$, act essentially to dilute the magnetic interactions
in the system. Distinct layers of this magnetic lattice are in a AB
stacking and interlayer interaction may be possible, however, consecutive
layers are separated by $\approx9.8\text{ }\mathring{\text{A}}$ and
this interaction is likely weak. 

\begin{figure}
\begin{centering}
\includegraphics[scale=0.3]{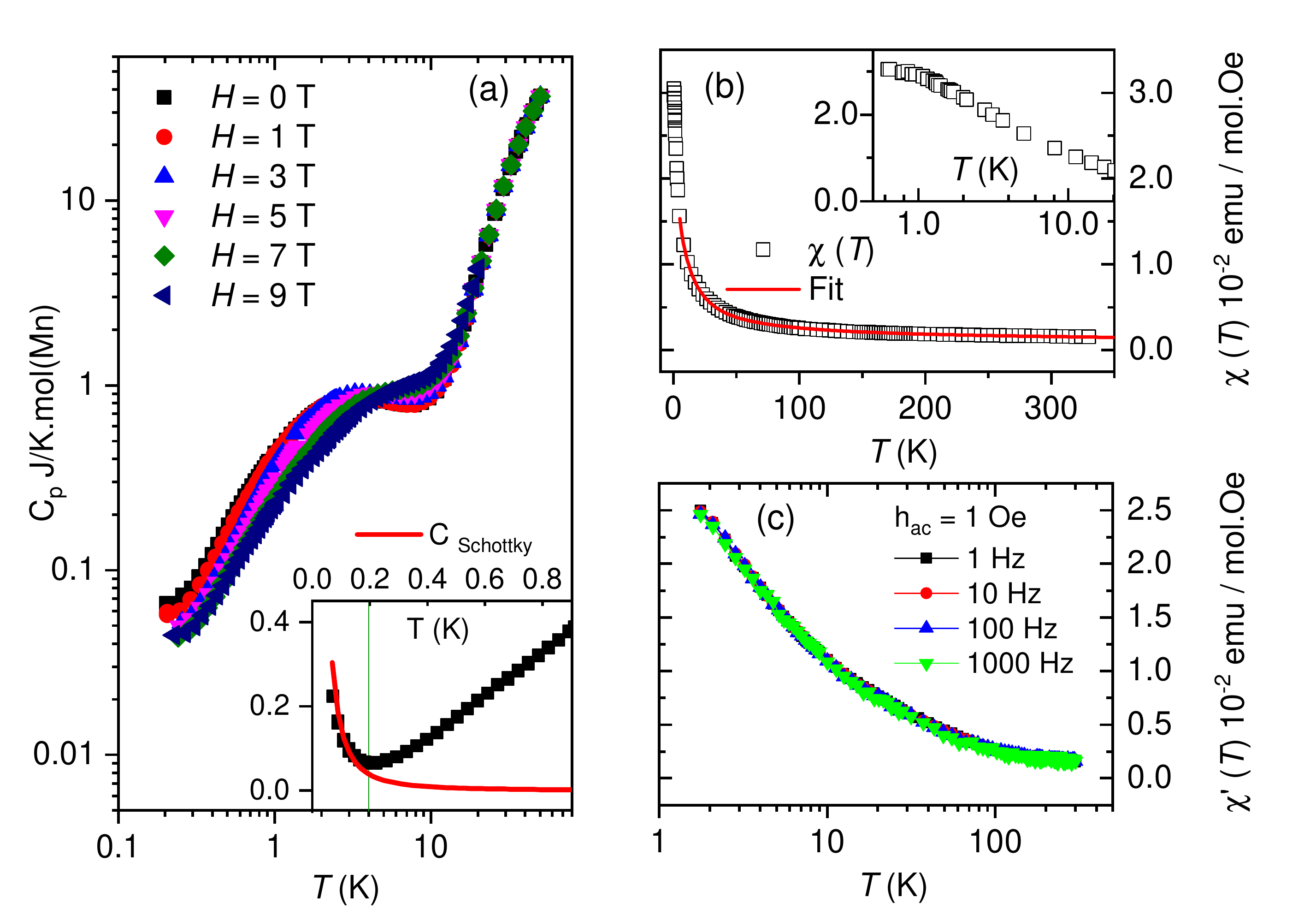}
\par\end{centering}
\caption{$(a)$ Heat capacity $(C_{p})$ measurements as a function of $T$
and applied field $H$. The inset shows the Mn nuclear contribution
to the heat capacity. $(b)$ DC Susceptibility {[}$\chi(T)${]} measurements
as a function of $T$ ( $0.6\protect\leq T\protect\leq330$ K). The
inset shows $\chi(T)$ data for $T<2$ K. The fitting (thick line)
is discussed in Ref. \cite{Cantarino_supplemental}. $(c)$ AC Susceptibility
{[}$\chi'(T)${]} measurements as a function of $T$ ( $1.8\protect\leq T\protect\leq300$
K) at distinct frequencies. \label{fig:general}}
\end{figure}

In panel $(a)$ of Fig. \ref{fig:general} we present heat capacity
measurements, $C_{p}$, as a function of $T$ ($0.2<T<100$ K) for
distinct values of $H$. The data give no evidence for a phase transition
down to $T=0.2$ K , presenting only a broad anomaly peaking at about
$3$ K for $H=0$, that we shall interpret as a crossover to the dynamic
magnetic state. Below and about $T\lesssim0.2$ K, the data present
an upturn at low $-T$ that could be interpreted as the tail of a
transition at further lower temperatures. This upturn, however, is
actually due to the Mn nuclear contribution to the heat capacity.
This conclusion is supported by the data presented in the inset of
the same figure, wherein we compare the $H=0$ data with the Mn nuclear
heat capacity as measured in experiments with MnNi \cite{proctor_nuclear_1967}.
It is shown that it scales well with our data and, therefore, it is
safe to assert that no phase transition is observed down to $T\approx0.1$
K. 

In panel $(b)$, $\chi(T)$ measurements are presented in the interval
$0.6<T<330$ K. To describe the data in the range $T>5.5$ K, which
is above the crossover region, we refer to our model \cite{Cantarino_supplemental}
including magnetic dimers and trimers plus orphan spins. The obtained
parameters associated to dimers and trimers are $J_{1}=176(5)$ K
and $J_{2}=0.59(5)J_{1}$. In this parameter range, the trimer ground
state is formed by effective $S=1/2$ spins, which, along with the
$S=3/2$ spins from the Mn$^{4+}$ orphans, are the remaining magnetic
degrees of freedom at low$-T$. In the inset, it can be observed a
trend towards saturation which cannot be related to a simple Curie-Weiss
behavior. It rather indicates the absence of free paramagnetic spins
in the system and thus we conclude that the effective $S=1/2$ spins
and orphan spins are correlated at low-$T$. 

To estimate the  energy scale of the correlated state of trimers and
orphans, we model the orphan-trimer subsystem by a Curie-Weiss susceptibility,
with parameters $\theta_{\text{eff}}$ and $C_{\text{eff}}$,  obtaining
$\theta_{\text{eff}}=-7.5(5)$ K and $C_{\text{eff}}=0.19(1)$ emu.K/mol
(f.u.) \cite{Cantarino_supplemental}. The value of $\theta_{\text{eff}}$
supports the idea of a highly frustrated system, with a lower bound
for the frustration parameter of $f\gtrsim7.5/0.1=75$. Here, $\theta_{\text{eff}}$
is to be understood as a phenomenological constant which sets the
energy scale of the interactions between the remaining magnetic degrees
of freedom: orphan-orphan, orphan-trimer and trimer-trimer interactions.
In the context of the discussion of Fig. \ref{fig:Structure}$(c)$,
it reflects mainly the intralayer interactions of the diluted magnetic
lattice. As for $C_{\text{eff}}=0.19(1)$ emu.K/mol (f.u.), it compares
relatively well with $C\approx0.14$ emu.K/mol (f.u.) expected for
$1/16$ mols of Mn $S=3/2$ plus effective $S=1/2$ spins per formula
unit of BaTi$_{1/2}$Mn$_{1/2}$O$_{3}$. In Fig. \ref{fig:general}$(c)$,
$\chi'(T)$ measurements ($1.8<T<300$ K) at a broad range of frequencies
($\nu$) do not depend on $\nu$ even in the interval $1.8-3$ K,
that is below the crossover region. Such dependency could be expected,
for instance, in spin glass systems. 

\begin{figure}
\begin{centering}
\includegraphics[scale=0.28]{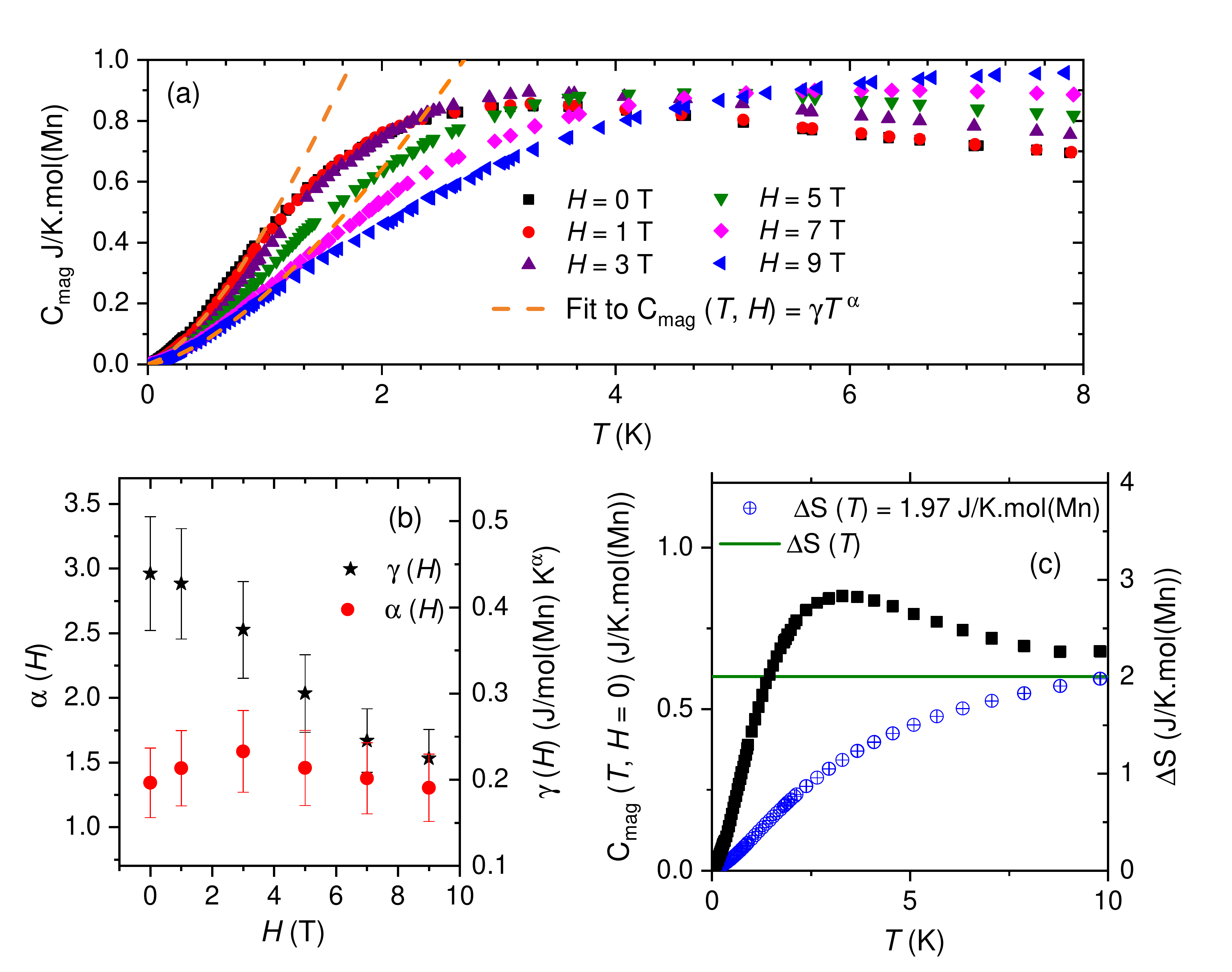}
\par\end{centering}
\caption{$(a)$ Magnetic specific heat, $C_{\mbox{mag}}(T)$, obtained after
the subtraction of the lattice contribution. The dashed lines are
representative fittings ($H=0$ and $H=9$ T) to $C_{\mbox{mag}}(T)=\gamma(H)T^{\alpha(H)}$
. $(b)$ Full set of field dependent parameters $\gamma(H)$ and $\alpha(H)$
obtained from this fitting .$(c)$ Zero field magnetic specific heat
data and the total magnetic entropy $\Delta S(T)$. \label{fig:Heatcapacity}}
\end{figure}

The nature of the dynamic magnetic state will manifest in the behavior
of magnetic specific heat $C_{\mbox{mag}}$. Following the subtraction
of the phonon contribution ($C_{\mbox{lattice}}$, \cite{Cantarino_supplemental}
), $C_{\mbox{mag}}$ is presented in panel $(a)$ of Fig. \ref{fig:Heatcapacity}
for $H$ up to $9$ T. Inspecting the broad anomalies, we observe
that the peaks displace to higher temperatures (from $T\approx3-7$
K) with increasing field. To characterize the magnetic excitations
of the low-$T$ phase, we focus on the low temperature region ($0.2\leq T\leq1.0$
K). It is adopted that $C_{\mbox{mag}}(T)$ will in general follow
a power law in the low$-T$ region and the data are fitted to the
expression $C_{\mbox{mag}}(T)=\gamma(H)T^{\alpha(H)}$ . The dashed
lines are representative ($H=0$ T and $H=9$ T) fittings. The resulting
parameters, $\gamma(H)$ and $\alpha(H)$, are shown in Fig. \ref{fig:Heatcapacity}$(b)$.
Within error bars, $\alpha(H)$ is nearly constant assuming a value
of about $\approx1.45(5)$, suggesting gapless excitations of this
low-$T$ dynamic magnetic phase. Moreover, $C_{\mbox{mag}}/T$ do
not diverge as $T\rightarrow0$, meaning that our system is away from
quantum criticality \cite{singh_spin_2013}. The value of $\gamma(H)$
decreases with increasing field from $\gamma\approx0.45$ to $\gamma\approx0.25$
J/mol(Mn)K$^{\alpha}$ at $H=9$ T, indicating quenching of the magnetic
entropy as a function of the field, albeit there are no free spins
in the system at this temperature range. These values are relatively
large when compared to spin liquid candidates \cite{li_gapless_2015}.

The nature of the zero field $C_{\mbox{mag}}$, in particular, is
key to estimate the total entropy recovered in the ground state and
to reveal the total fraction of spins that are part of the proposed
correlated state. In Fig. \ref{fig:Heatcapacity}$(c),$ we show the
heat capacity data and the magnetic entropy {[}$\Delta S(T)${]} for
$H=0$ in the interval $0.1\leq T\leq10$ K, within which the entropy
contribution from excited states of the trimer is negligible. The
total entropy recovered by the system amounts to $\approx1.97(5)$
J/K.mol(Mn). This number must be compared with the entropy expected
from the remaining magnetic degrees of freedom at low$-T$ \cite{Cantarino_supplemental},
that amounts to $1/8$ mols of $S=3/2$ and effective $S=1/2$ spins
per mol of Mn atoms. Thus, the total expected entropy is $R/8(\ln4+\ln2)\approx2.16$
J/K.mol(Mn). The experimental result is certainly larger than what
is expected for the orphan spins only {[}$\approx1.44$ J/K.mol(Mn){]},
further suggesting that the remaining magnetic degrees of freedom
must include $S=3/2$ and effective $S=1/2$ spins. Moreover, the
trimers cannot be in a simple paramagnetic state for such state would
imply (i) sizable Schottky anomalies in the field dependent heat capacity
measurements, and (ii) a Curie-like response in the magnetic susceptibility,
both which are not observed \cite{Cantarino_supplemental}. Overall,
if we consider the total entropy due to trimers and orphans in the
system, we may assert that we recover a total of $\approx90\%$ of
the expected system entropy. 

\subsection{$\mu$SR ZF and LF experiments}

\begin{figure}[t]
\begin{centering}
\includegraphics[scale=0.3]{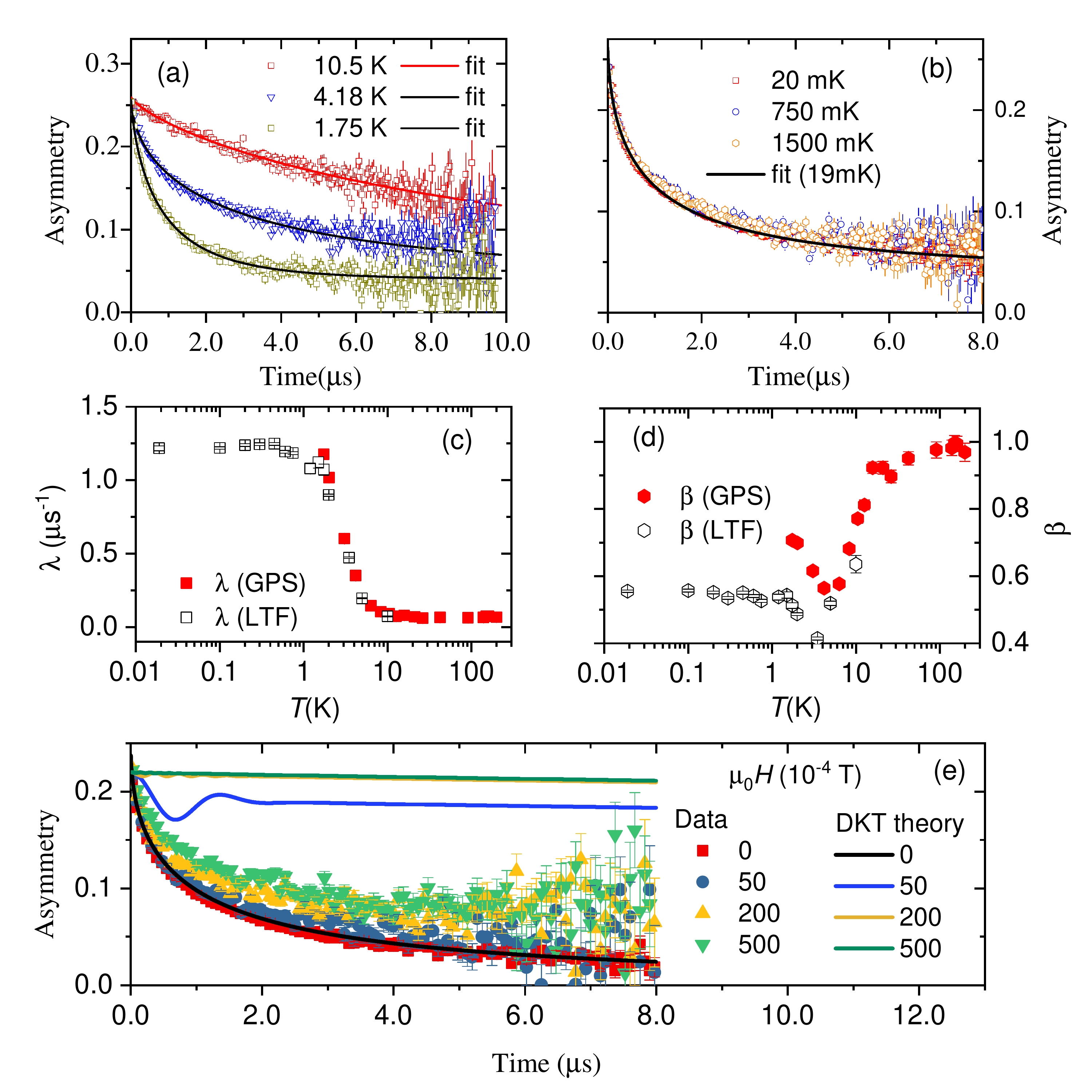}
\par\end{centering}
\caption{$(a)-(b)$ Representative ZF$-\mu$SR spectra measured at the $(a)$
crossover region and $(b)$ in the low$-T$ region. Thick lines are
the fittings of the data to a stretched exponential. Temperature dependency
of the $(c)$ $\mu^{+}$ spin relaxation rate, $\lambda$ and $(d)$
stretching exponent, $\beta$ as a function of $T$ . $(e)$ LF experiment
($T=0.1$ K) spectra and comparison with the DKT theory.\label{fig:muons}}
\end{figure}

We now turn our attention to our $\mu$SR experiments \cite{blundell_spin-polarized_1999}.
In Fig. \ref{fig:muons}$(a)$-$(d)$ we present ZF $\mu$SR data
measured for temperatures down to $T=0.019$ K. Down to the lowest
temperatures, the spectra (Figs. \ref{fig:muons}$(a)$-$(b)$) can
be described quite well by an exponential decay, and the data do not
present the characteristic damped oscillations of a magnetic ordered
phase \cite{blundell_spin-polarized_1999}. Fig \ref{fig:muons}$(a)$
shows the temperature interval for which the spectra change the most
($1.5<T<10$ K), characterizing a crossover behavior in this region.
Fig \ref{fig:muons}$(b)$ display the low temperature spectra, which
can be almost described by a single fitting below $T<1.5$ K. The
theoretical curves represent the best fitting to a stretched exponential
which describes the asymmetry function $A(t)=A_{0}\exp-(\lambda t)^{\beta}+B$,
where $A_{0}$ is the starting asymmetry, $B$ describes the constant
background mostly due to muons that did not stop in the sample and
$\lambda$ and $\beta$ are, respectively the relaxation rate and
the stretching exponent. The muon environment was investigated by
means of DFT calculations. The result proposes that the muon is located
close to the Oxygen between the two Barium sites \cite{Cantarino_supplemental}.

The resulting fitting parameters $\lambda$ and $\beta$ are shown
in Figs \ref{fig:muons}$(c)$-$(d)$. For $T>10$ K, $\lambda$ is
nearly temperature independent, as typical for simple paramagnets.
In the $T-$ region $1.5<T<10$ K, $\lambda$ goes through a steep
increase from $\approx0.06\mu\text{s}^{-1}$ up to a constant value
$\approx1.2\mu\text{s}^{-1}$ for $T<1.5$ K, characteristic of a
crossover behavior. The $\beta$ coefficient do also display a crossover
behavior in this same $T-$ region . Its value decreases from $\approx1$
at high $T$ to a constant value close to $0.5$ for $T<1.5$ K. This
value is well above $1/3$ that would be expected for glassy dynamics
of simple systems, indicating the absence of static local fields \cite{uemura_muon_1985,blundell_spin-polarized_1999}.
It is noteworthy that the crossover region detected by the $\mu$SR
corresponds closely to the region where the broad peak is observed
for the heat capacity measurements. Our data illustrate a good qualitative
agreement between macroscopic and microscopic measurements. 

We have as well performed LF measurements at $T=0.1$ K and the results
are presented in Fig \ref{fig:muons}$(e)$. The data is compared
with the expected results from the dynamical Kubo-Toyabe (DKT) theory.
The change in the spectra as a function of field (from $0$ to $0.05$
T) is much less dramatic than the expected from the DKT theory. This
is a compelling piece of evidence for dynamic magnetism at temperatures
down to $T=0.1$ K. Following Refs. \cite{uemura_spin_1994,li_muon_2016},
the spin fluctuation rates for temperatures above and below the crossover
can be estimated \cite{Cantarino_supplemental}. The results are $\nu_{T>10\text{ K}}=7.4(1)\times10^{10}$
Hz and $\nu_{T<1.5\text{ K}}=3.7(6)\times10^{6}$ Hz, suggesting,
without symmetry breaking, the onset of long-time spin correlations
in the system for $T<1.5$ K.

\subsection{The nature of the dynamic magnetic state}

By direct inspection, the ZF spectra at low $T$ in Fig.\ref{fig:muons}$(b)$
lack the characteristic recovery of $1/3$ of the $\mu^{+}$ polarization
expected for glassy systems \cite{uemura_muon_1985}, thus we start
discussing the possibility of a spin liquid state. The experimental
results from heat capacity are compatible with a gapless spin liquid.
Because the magnetic susceptibility saturates at low $T$, a strong
candidate would be a spin liquid with a spinon Fermi-surface. However,
for the latter, we expect $C_{\mbox{mag}}\sim T^{2/3}$ \cite{li_muon_2016,motrunich05}
at zero field, unlike the $\approx T^{1.4-1.5}$ we observe. Another
possibility would be an $U(1)$ Dirac spin-liquid \cite{ran07,ryu07,he17}.
Here, one expects $C_{\mbox{mag}}\sim T^{2}$ at zero fields and $C_{\mbox{mag}}\sim T$
at low-fields, and that $C_{\mbox{mag}}$ should increase with the
field, which we also do not observe. 

A similar drawback is recently under debate in YbMgGaO$_{4}$, where
the observed spin liquid behavior is not captured by the minimal theoretical
model \cite{li_muon_2016,li_rare-earth_2015,li_gapless_2015,paddison_continuous_2017}.
One possible scenario to understand the reported experimental behavior
is to take into account the presence of disorder in the system \cite{li_crystalline_2017},
which may have non-trivial effects, even leading to the destruction
of long-range magnetic order \cite{zhu17,andrade18,parker18}. Since
the $\mu$SR results for YbMgGaO$_{4}$ are qualitatively similar
to ours \cite{li_muon_2016}, disorder is likely a relevant parameter
to understand the behavior of BaTi$_{1/2}$Mn$_{1/2}$O$_{3}$. 

In this line of thinking, the large jump in $\lambda$ is a drawback
to the interpretation in terms of a spin liquid state, since it is
reminiscent of spin glass behavior \cite{uemura_muon_1985,uemura_spin_1994,keren_probing_1996,tran_low-temperature_2018}.
Furthermore, a spin glass state is favored by the presence of a disordered
magnetic lattice of $S=3/2$ and effective $S=1/2$ spins as suggested
by our structural analysis. However, large jumps of $\lambda$ can
be found as well for PbCuTe$_{2}$O$_{6}$ \cite{khuntia_spin_2016}
or Tm$_{3}$Sb$_{3}$Zn$_{2}$O$_{14}$ \cite{ding_possible_2018},
which are all recently proposed spin liquid candidates, presenting
a characteristic large $\lambda$ of about $\approx1.0\mu\text{s}^{-1}$
at low-$T$, as in the present case. The steep increase in $\lambda$
suggests a glass transition temperature ($T_{g}$) of about $T_{\text{g}}\approx1-2$
K and this must be contrasted with the frequency independence of $\chi'(T)$,
specially in the region $1.8-3$ K. Certainly, $\chi'(T)$ measurements
down to lower temperatures are needed for a more conclusive statement.

Based upon the heat capacity analysis, we can account for $90\%$
of the total entropy of the system. This result almost exclude the
possibility that the probed dynamic magnetic state could relate to
a crossover region to an ordered state at further lower temperature.
In particular, ferrimagnetic order due to the antiferromagnetic coupling
of the unlike $S=3/2$ and $S=1/2$ spins \cite{lieb62,iqbal_quantum_2018}
is not to be expected. The existence of residual entropy is compatible
with a spin glass and the dynamic magnetism probed by the ZF spectra
at $0.019$ K could be due to the fraction of spins that are not frozen
due a distribution of time scales in a spin glass. However, residual
entropy is also observed in spin liquid candidates \cite{kumar_$mathrmba_3m_xmathrmti_3ensuremath-xmathrmo_9$_2016,kumar_$mathrmsc_2mathrmga_2mathrmcuo_7$:_2015}.
In short, the low-$T$ dynamic magnetic state in our system displays
spin liquid behavior, although a spin glass state, induced by the
disordered lattice of orphans and trimers \cite{zhu17,andrade18,parker18},
is a strong candidate to interpret our results. 

\section{Summary and conclusions}

We have performed macroscopic and microscopic experiments in the disordered
hexagonal double perovskite BaTi$_{1/2}$Mn$_{1/2}$O$_{3}$. This
complex material displays a rich magnetic behavior with the physics
at high temperatures dominated by the presence of trimers, dimers,
and orphan spins \cite{garcia_magnetic_2015}. At lower temperatures,
the effective magnetic degrees of freedom, composed by orphan spins
and trimers, are found to be correlated but no phase transition is
detected down to $T=0.1$ K, despite the effective exchange couplings
being of the order of $10$ K. $\mu$SR measurements then show substantial
evidence for dynamic magnetism down to $T=0.019$ K, the nature of
which was discussed in terms of a spin liquid and a spin glass scenario.
The proper characterization of this state requires further theoretical
and experimental efforts.
\begin{acknowledgments}
MRC acknowledges CNPq grant No.131117/2017-3 for financial support.
The authors acknowledge CNPEM-LNLS for the concession of beam time
(proposal No. $20160245$). The XRD$1$ beamline staff is acknowledged
for the assistance during the experiments. RSF acknowledges FAPESP
grant No.2015/16191-5 and CNPq grant No.306614/2015-4. RLS acknowledges
FAPEMIG-MG (APQ-02256-12) and CAPES Foundation (Brazil) for grant
EST-SENIOR-88881.119768/2016-01. Author VG is supported by GR 4667.
RS and HHK are partially supported by DFG SFB 1143 for the project
C02. ECA acknowledges FAPESP Grant No. 2013/00681-8 and CNPq Grant
No. 302065/2016-4.
\end{acknowledgments}

\bibliographystyle{apsrev4-1}
\bibliography{SLiquid_references}

\end{document}